\documentclass[conference]{IEEEtran}
\IEEEoverridecommandlockouts
\usepackage{cite}
\usepackage{amsmath,amssymb,amsfonts}
\usepackage{algorithm}
\usepackage{algpseudocode}
\usepackage{graphicx}
\usepackage{textcomp}
\usepackage{xcolor}
\usepackage{comment}
\usepackage{fancyhdr}
\usepackage{eso-pic}

\usepackage{atbegshi} 

\def\BibTeX{{\rm B\kern-.05em{\sc i\kern-.025em b}\kern-.08em
    T\kern-.1667em\lower.7ex\hbox{E}\kern-.125emX}}

\newlength{\tightskip}
\expandafter\def\expandafter\normalsize\expandafter{%
    \normalsize
    \setlength\abovedisplayskip{\tightskip}
    \setlength\belowdisplayskip{\tightskip}
    \setlength\abovedisplayshortskip{\tightskip}
    \setlength\belowdisplayshortskip{\tightskip}
}

\fancypagestyle{firstpage}{
  \fancyhf{} 
  \fancyfoot[C]{\footnotesize
    \begin{minipage}{\textwidth}
    \centering
    \color{red}{
© 2026 IEEE. This paper has been accepted for presentation at the 2026 IEEE Power \& Energy Society General Meeting (PESGM).}
    \end{minipage}
  }

}

\begin{document}

\title{Communication Network-Aware Missing Data Recovery for Enhanced Distribution Grid Visibility\\
\thanks{This work is supported by U.S. Department of Energy Office of Cybersecurity, Energy Security, and Emergency Response under award CR000039.}
}

\author{
\IEEEauthorblockN{
Biswas Rudra Jyoti Arka\textsuperscript{1},
Md. Zahidul Islam\textsuperscript{1},
Yuzhang Lin\textsuperscript{2},
Vinod M. Vokkarane\textsuperscript{3},
Junbo Zhao\textsuperscript{4}
}
\IEEEauthorblockA{\textsuperscript{1} School of Electrical, Computer, and Biomedical Engineering, Southern Illinois University Carbondale, IL 62901, USA}
\IEEEauthorblockA{\textsuperscript{2}Department of Electrical and Computer Engineering, New York University, New York, NY 10012, USA}
\IEEEauthorblockA{\textsuperscript{3}Department of Electrical and Computer Engineering, University of Massachusetts Lowell, MA 01854, USA}
\IEEEauthorblockA{\textsuperscript{4}Thayer School of Engineering, Dartmouth College, Hannover, NH 03755, USA}

Email: mdzahidul.islam@siu.edu
}

\maketitle
\thispagestyle{firstpage}
\begin{abstract}
Power distribution systems increasingly rely on dense sensor networks for real-time monitoring, yet unreliable communication links and equipment malfunctions often result in missing or incomplete measurement sets at the operating center, requiring accurate data recovery techniques. Most existing approaches operate solely on the available measurements and overlook the role of the communication network that delivers sensor data, leading to large, spatially correlated losses when multiple sensors share failing communication links. This paper proposes a communication-aware framework that integrates routing constraints with low-rank matrix completion to improve data recovery accuracy under communication failures. Sensors are grouped into balanced clusters, and routing paths are designed to limit intracluster sensors sharing a common communication path, preventing complete data loss within any cluster. The remaining measurements for each cluster are then recovered using an optimal singular value thresholding (OSVT) method. Simulation results on the IEEE standard test feeder with real-world data demonstrate that the proposed framework significantly improves recovery accuracy compared to communication-agnostic, measurement-only methods.

\end{abstract}

\section{Introduction}

As the integration of distributed energy resources (DERs) and data centers increases, power distribution systems are experiencing a paradigm shift in their operation. Traditional, historical pattern-based practices are being replaced by real-time, data-driven decision-making. Different types of sensors are being deployed, and communication networks are being upgraded to collect the large volume of sensor measurements needed to enable grid situational awareness \cite{b2}. However, unreliable communication links and equipment malfunctions can lead to missing data and incomplete measurement sets at the operating center, hindering operational decision-making and control actions \cite{b3}.

Several existing frameworks have attempted to mitigate the effects of measurement loss and communication interruptions in power systems through redundancy and data-driven measurement recovery techniques. An adaptive SVT-based algorithm incorporating MMSE estimation and SURE optimization was proposed in \cite{b8} to handle non-uniform missing data and noise in distribution grids. In \cite{b9}, the authors evaluated matrix completion-based recovery methods and demonstrated that matrix completion can outperform other techniques when observations are limited or source statistics are imperfect. The authors in \cite{b10} proposed low-rank matrix completion methods for recovering synchrophasor measurement data, with an emphasis on correlated erasures. An adaptive phasor measurement unit (PMU) data recovery approach was introduced in \cite{b11}, which first identifies the type of missing data and then employs different strategies to recover ambient versus disturbance data, thereby improving accuracy.
To emphasize real-time performance, a lightweight data-driven filter based on low-rank matrix recovery was developed in \cite{b12} for real-time PMU data reconstruction.
To improve scalability, Ref.~\cite{b13} employed a network embedding strategy that constructs a spatio-temporal graph of large systems and reduces the associated features into a lower-dimensional representation for use in a regression-based neural network recovery framework. Finally, the authors of \cite{b1} implemented an optimal singular value thresholding (OSVT) method for noisy PMU data, capturing the spatio-temporal dependencies among sensors.

Despite substantial progress in measurement recovery, existing approaches exhibit key limitations. Current methods either recover each sensor individually using historical patterns or reconstruct all measurements concurrently from the subset of sensors whose data become available at the operating center; however, both strategies struggle to scale to large distribution systems. More critically, existing work overlooks the role of the communication network, treating it merely as a source of missing data rather than as part of the recovery formulation. As a result, sensors essential for maintaining low-rank structure, and therefore high recovery accuracy, may become simultaneously unavailable if they share a common communication link that fails. Incorporating communication-aware constraints and cluster-wise recovery becomes essential to preserving low-rank structure, ensuring sufficient data availability, and improving overall recovery performance.

To address these limitations, this paper proposes a communication-aware recovery framework that embeds communication network routing directly into the missing data recovery process. \textit{The central idea is to define routing paths such that only a specified percentage of sensors within a recovery group are allowed to share a common communication link.} This ensures that no group of sensors loses all of its measurements simultaneously due to one or multiple link failures, thereby preserving sufficient data at the operating center to support accurate recovery. To realize this idea, sensors are first clustered based on their historical measurements. Cluster-wise communication routes are then identified, which guarantees that each cluster retains enough available data during link failures and preserves the low-rank structure required for effective matrix completion. Finally, a matrix completion method based on optimal singular value thresholding (OSVT) is employed to reconstruct missing sensor measurements, such as voltages and nodal power injections, with high accuracy.

\section{Problem Statement}

Let $\mathbf{X}^k \in \mathbb{R}^{T \times M}$ denote the measurement matrix (e.g., voltages or power injections) acquired over a time horizon of length $T$ from $M$ sensors belonging to cluster $k$. During an event, the failure of one or multiple communication links causes partial data loss across clusters, resulting in an incomplete observation matrix $\mathbf{X}^k_{\Omega}$, where $\Omega \subseteq \{1, \dots, T\} \times \{1, \dots, M\}$ denotes the set of available entries.

The objective of this study is to recover the missing data and reconstruct a complete observation matrix $\mathbf{\hat{X}}^k$ that closely approximates the original measurements by leveraging the available entries and the spatial correlations encoded in the grid topology $\mathcal{G}_p$. This problem can be formulated as:
\begin{equation}
\min_{\mathbf{\hat{X}}^k} \; \| P_{\Omega}(\mathbf{\hat{X}}^k - \mathbf{X}^k) \|_F^2
\quad \text{s.t.} \quad \mathbf{\hat{X}}^k \in \mathcal{M}(\mathcal{G}_p),
\end{equation}
where $P_{\Omega}(\cdot)$ is a projection operator that preserves the observed entries, and $\mathcal{M}(\mathcal{G}_p)$ encodes (implicitly via clustering and routing) the spatial correlations among sensors derived from the grid topology. 

The solution $\mathbf{\hat{X}}^k$ represents the recovered full measurement matrix that remains consistent with the underlying physical constraints, ensuring resilient grid monitoring under communication failures.

\vspace{-5pt}
\section{Methodology}

\begin{figure}
    \centering
    \includegraphics[width=1\linewidth]{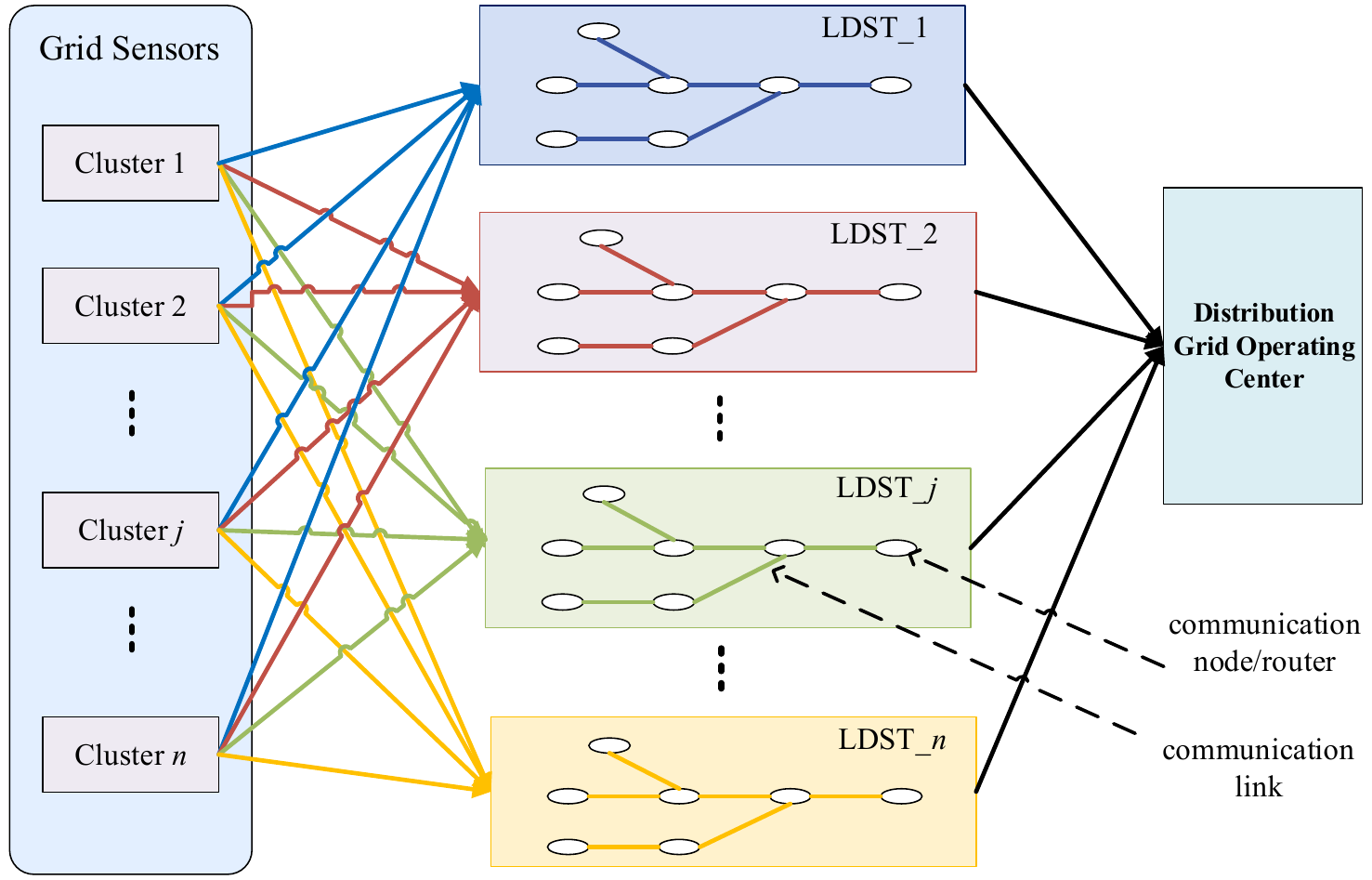}
    \caption{Conceptual diagram of the proposed framework.}
    \label{fig:concept}
\end{figure}

An overview of the proposed framework is shown in Fig.~\ref{fig:concept}. Sensors are first grouped into $n$ clusters based on their historical measurements, so that sensors with similar profiles belong to the same cluster. Starting from the operating center, a set of link-disjoint Steiner trees (LDSTs) is then constructed in the communication network to connect the sensors in different clusters according to operator-defined criteria. These LDSTs are mutually link-disjoint, meaning that a failure on one communication link does not affect the delivery of data carried by other LDSTs. In the figure, each LDST is illustrated using a distinct color. As shown, even if a communication link fails, some sensors within each cluster can still report data because they are routed through different LDSTs. Finally, at the operating center, OSVT is applied to each cluster to recover missing sensor data with improved accuracy. The detailed methodology is presented in the following subsections.

\subsection{Constrained \textit{k}-Means Clustering}

A constrained \textit{k}-means clustering method is applied to group the sensors based on their historical measurement profiles. Let $\mathbf{z}_i$ denote the feature vector extracted from the historical measurements of sensor~$i$. In this work, the feature vector consists of the mean, maximum, minimum, median, and variance of the historical data. The clustering objective is to partition the sensors into $K$ clusters by minimizing the standard \textit{k}-means cost \cite{b4}:
\begin{equation}
\min_{\{C_k,\;\boldsymbol{\mu}_k\}}
\sum_{k=1}^{K} \sum_{\mathbf{z}_i \in C_k}
\|\mathbf{z}_i - \boldsymbol{\mu}_k\|^{2},
\end{equation}
where $\boldsymbol{\mu}_k$ is the centroid of cluster $C_k$.

Although the objective remains identical to classical \textit{k}-means, additional constraints are introduced to better support communication-aware recovery. Each cluster is required to be of almost equal size, with its cardinality maintained close to $(\#\text{sensors})/K$ \cite{islam2025smart}. These size constraints prevent clusters from becoming too small, which would limit the amount of data available for low-rank recovery, or too large, which would reduce intra-cluster similarity. By enforcing balanced cluster sizes, each cluster retains sufficient and coherent measurements to enable effective cluster-wise data reconstruction under communication link failures.

\subsection{Communication Route Formation for Clusters} \label{sec:LDST}

After forming the sensor clusters, communication routes are established to deliver measurements to the distribution management system (DMS). Conventional routing practices typically ignore grid-side knowledge, as communication paths are established without considering the electrical relationships among sensors. To incorporate grid awareness, we construct a set of LDSTs in the communication network, denoted by:
\begin{equation}
\mathcal{T} = {T_1, T_2, \ldots, T_M},
\end{equation}
where each $T_m$ forms an independent communication subnetwork satisfying:
\begin{equation}
E(T_m) \cap E(T_n) = \emptyset, \qquad \forall m \neq n.
\end{equation}
This ensures that a link failure affecting one LDST does not compromise sensors routed through the others.

A central routing requirement is how sensors within a cluster $C_k$ are distributed across LDSTs. Let $|C_k|$ denote the cluster size, and let $C_k^{(m)} \subseteq C_k$ be the subset of sensors assigned to LDST $T_m$. To prevent complete loss of cluster data under link failures, we impose the constraint:
\begin{equation} \label{eq:LDST}
\frac{|C_k^{(m)}|}{|C_k|} \le \alpha_k^{\text{LDST}}, \qquad \forall m,
\end{equation}
where $\alpha_k^{\text{LDST}} \in (0,1)$ is a tunable parameter. For example, if $|C_k| = 10$ and $\alpha_k^{\text{LDST}} = 0.3$, at most three sensors from cluster $k$ may be routed through the same LDST. Without this constraint, intra-cluster sensors, often physically co-located and exhibiting highly similar measurements, would naturally tend to share the same LDST, increasing the likelihood of complete cluster-level data loss during communication disruptions.

In addition to this grid-aware constraint, we also enforce communication requirements such as link bandwidth and load balancing across the LDSTs. The complete formulation of LDST construction and communication load balancing is discussed in our previous work in~\cite{islam2024observability}. While assigning backup LDSTs to each cluster improves resilience against single-link failures, creating many LDSTs is impractical due to the sparse communication topology of distribution networks. Furthermore, multiple simultaneous link failures can still result in missing data at the DMS. To address this limitation, we discuss missing data recovery using the LDST structure in the next subsection.

\begin{algorithm}[t]
\caption{Optimal Singular Value Thresholding (OSVT)}
\begin{algorithmic}[1]
\Require Observation matrix $X \in \mathbb{R}^{m \times n}$ with missing or noisy entries
\Ensure Reconstructed low-rank estimate $\widehat{M}$

\State \textbf{Normalize the data:} Scale $X$ into $[-1, 1]$ as
\begin{equation}
y_{ij} = \frac{x_{ij} - 0.5(a + b)}{0.5(b - a)},
\end{equation}
where $a = \min(X)$ and $b = \max(X)$. Let the scaled matrix be $Y$.

\State \textbf{Singular Value Decomposition (SVD):}
\begin{equation}
Y = \sum_{i=1}^{m} \sigma_i u_i v_i^{T},
\end{equation}
where $\sigma_i$ are singular values and $u_i, v_i$ are the left and right singular vectors.

\State \textbf{Thresholding:}  
Select the set of retained singular values
\begin{equation}
S = \{ \sigma_i : \sigma_i > \sigma_{\text{th}} \},
\end{equation}
where the optimal threshold $\sigma_{\text{th}}$ is
\begin{equation}
\sigma_{\text{th}} = \sqrt{2(\zeta + 1) + \frac{8\zeta}{(\zeta + 1) + \sqrt{\zeta^2 + 14\zeta + 1}}},
\end{equation}
and $\zeta = m/n$.

\State \textbf{Low-rank Reconstruction:}
\begin{equation}
\widehat{Y} = \sum_{\sigma_i \in S} \sigma_i u_i v_i^{T}.
\end{equation}

\State \textbf{Rescale to Original Range:}
Transform $\widehat{Y}$ back to the original data range $[a, b]$ to obtain
\begin{equation}
\widehat{M} = 0.5(b - a)\widehat{Y} + 0.5(a + b).
\end{equation}

\end{algorithmic}
\end{algorithm}

\subsection{Data Recovery Under Failures}

To capture the spatial and temporal relationships among sensor measurements within each cluster, the cluster datasets are first transformed into Page matrices~\cite{b7}. A Page matrix is constructed by arranging non-overlapping segments of a time-series as the columns of a matrix. Let $L$ denote the number of consecutive samples (rows) in each column segment, which defines the temporal window length, and the total time horizon determines how many such non-overlapping segments form the columns. This transformation has been shown to effectively reveal low-rank structure in real-world processes~\cite{b7}, thereby enabling reliable recovery of missing or corrupted data.

Once the cluster data are arranged into a Page matrix, low-rank matrix estimation is performed using OSVT. Several approaches exist for low-rank recovery, and Ref.~\cite{b1} demonstrated that OSVT offers improved performance for synchrophasor data recovery; therefore, it is adopted in this work. OSVT leverages the low-rank structure exposed by the Page matrix to estimate missing matrix entries, producing an accurate reconstruction of voltage and power-injection measurements under communication link failures. The steps of the OSVT procedure are summarized in Algorithm~1. The algorithm operates by first normalizing the observation matrix and applying singular value decomposition (SVD) to separate its dominant low-rank structure from missing entries. An optimal threshold is then applied to the singular values, retaining only those that correspond to meaningful signal components while discarding the rest. The retained components are recomposed to form a low-rank estimate, which is finally rescaled to the original data range to obtain the reconstructed measurement matrix.
\begin{figure*}
    \centering
    \includegraphics[height=.34\linewidth, width=.88\linewidth]{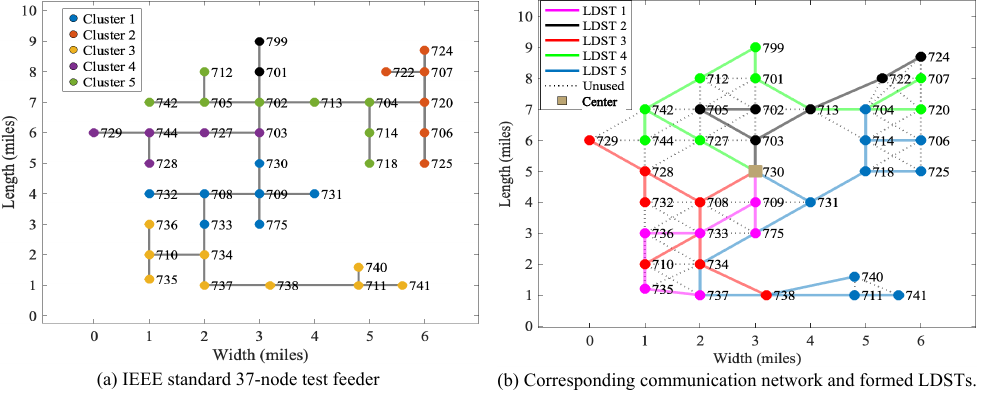}
    \caption{Test system used for performance evaluation: (a) IEEE 37-node test feeder with clusters based on nodal voltages; (b) Corresponding communication network and formed LDSTs.}
    \label{fig:test_system}
\end{figure*}
\vspace{-10pt}
\section{Case Study}

The proposed communication-aware data recovery framework is evaluated using the IEEE standard 37-node test feeder. A corresponding communication network is constructed by mapping the feeder onto a geographic region and allowing communication links between nodes located within a realistic distance threshold \cite{islam2024observability}. The resulting power and communication networks are shown in Fig.~\ref{fig:test_system}. Several distributed generators are also added to maintain acceptable voltage levels. Nodal power injections are derived by aggregating smart meter readings at each bus, where the London smart meter dataset is used to generate 30 days of time-series load data at 30-minute resolution, yielding 1440 samples~\cite{b6}. Subsequently, voltages and other nodal measurements are computed by solving the circuit using OpenDSS software.

To demonstrate the effectiveness of the proposed approach, multiple sensor types (e.g., voltage and power injection sensors) are assumed, with their measurements delivered via the communication network. Clusters are formed separately for each sensor type, and independent LDSTs are constructed for each cluster. Because each communication link is assumed to support multiple channels, multiple LDSTs can coexist on the same physical network by exploiting different channels \cite{islam2024observability}. The operating center is placed at bus~730 due to its central location in the network, and LDSTs are generated outward from this location along each adjacent link as detailed in Section~\ref{sec:LDST} and Ref.~\cite{islam2024observability}. Communication link failures are then simulated via Monte Carlo sampling to create missing data scenarios, and OSVT is applied to recover the missing measurements.


Two baseline methods are implemented for comparison. Baseline 1 mirrors the proposed pipeline but builds LDSTs without the intra-cluster routing constraint, allowing sensors in a cluster to share communication paths. Baseline 2 recovers intra-cluster data directly, without Page-matrix formulation or LDST constraints. All methods use identical missing-data scenarios and apply cluster-wise OSVT recovery. Performance is evaluated using MAE, RMSE, and MAPE~\cite{islam2025smart} to assess the benefits of communication-aware routing.

\begin{figure}
    \centering
    \includegraphics[width=0.48\textwidth]{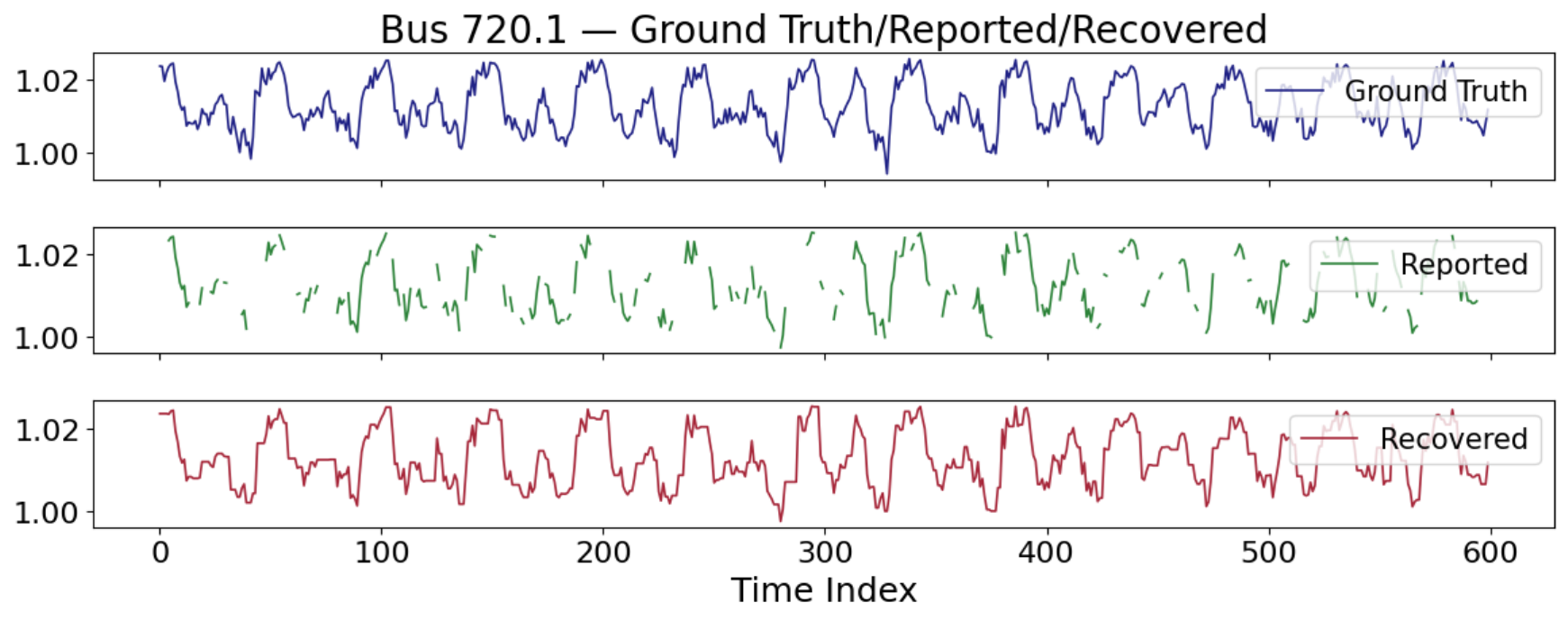}
    \caption{Comparison of ground-truth, reported, and recovered voltages for a single representative node.}
    \label{fig:1}
\end{figure}
\begin{figure}
    \centering
    \includegraphics[width=0.48\textwidth]{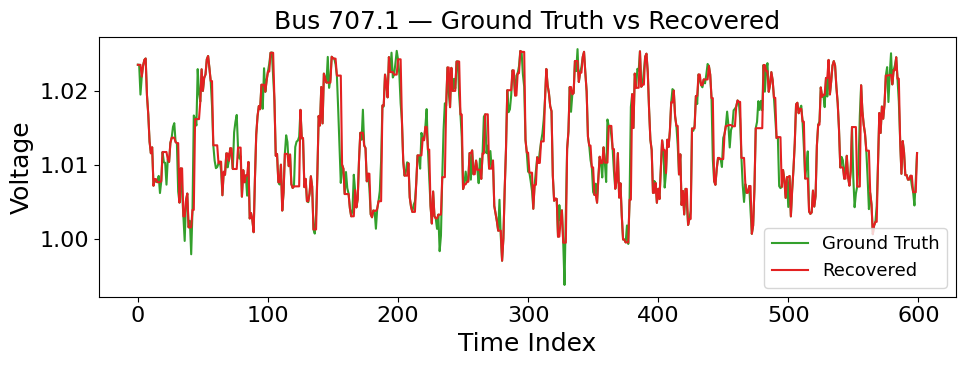}
    \caption{Ground truth versus recovered voltages for a single node.}
    \label{fig: volt overlap}
\end{figure}
\begin{table}
    \centering
    \caption{MAE comparison between proposed and baseline methods for voltage recovery under multiple link failures.}
    \label{tab:mae volt baseline}
    \begin{tabular}{ccccc}
        \hline
        \textbf{\textit{K}} & \textbf{Proposed} & \textbf{Baseline1} & \textbf{Baseline2} & Improvement (\%) \\
        \hline
        1        & 0.002568 & 0.002618 & 0.003092 & \textbf{1.83}  \\
        2        & 0.002450 & 0.002748 & 0.002859 & \textbf{10.84} \\
        3        & 0.003010 & 0.003255 & 0.003590 & \textbf{7.53}  \\
        4        & 0.002170 & 0.002453 & 0.002728 & \textbf{11.54} \\
        5        & 0.002296 & 0.002541 & 0.002731 & \textbf{9.88}  \\
        \hline
        \textbf{Combined} & \textbf{0.002580} & \textbf{0.002784} & \textbf{0.003044} & \textbf{7.33}  \\
        \hline
    \end{tabular}
\end{table}
\subsection{Performance Evaluation}

\textbf{1) Voltage Measurement Recovery.}
Performance is first evaluated for voltage measurements. It is assumed that voltage sensors are available at most nodes. For clustering, the voltage sensors are partitioned into five clusters using the constrained \textit{k}-means method, and LDSTs are subsequently formed for the voltage sensor nodes with $\alpha_k^{\text{LDST}} = 0.3$, as defined in Eq.~\eqref{eq:LDST}. The resulting voltage clusters and their associated LDSTs are shown in Fig.~\ref{fig:test_system}, which illustrates that intracluster sensors are routed through different LDSTs. For example, Cluster~2 (nodes 706, 707, 720, 722, 724, 725) is distributed across three LDSTs (LDSTs~2, 4, and 5), thereby maintaining cluster-level data availability even if multiple links in two LDSTs fail.

Missing data are generated using Monte Carlo simulation. At each time step, up to five random communication link failures are introduced to create extreme but realistic missing-data scenarios. The resulting incomplete measurements for each cluster are arranged into Page matrices, and OSVT is applied for recovery. The parameters $L$ and the time-window length associated with the Page matrix are tuned for best performance, with $L=8$ and six time windows yielding the most accurate voltage recovery. Fig.~\ref{fig:1} shows the ground-truth voltage, the reported data with missing entries, and the recovered measurements for a representative node. Overlaid ground-truth and recovered voltages for another node are also shown in Fig.~\ref{fig: volt overlap}. Both figures demonstrate that the proposed method reconstructs the voltage time series with high fidelity.

\begin{figure}[t]
    \centering
    \includegraphics[width=1\linewidth]{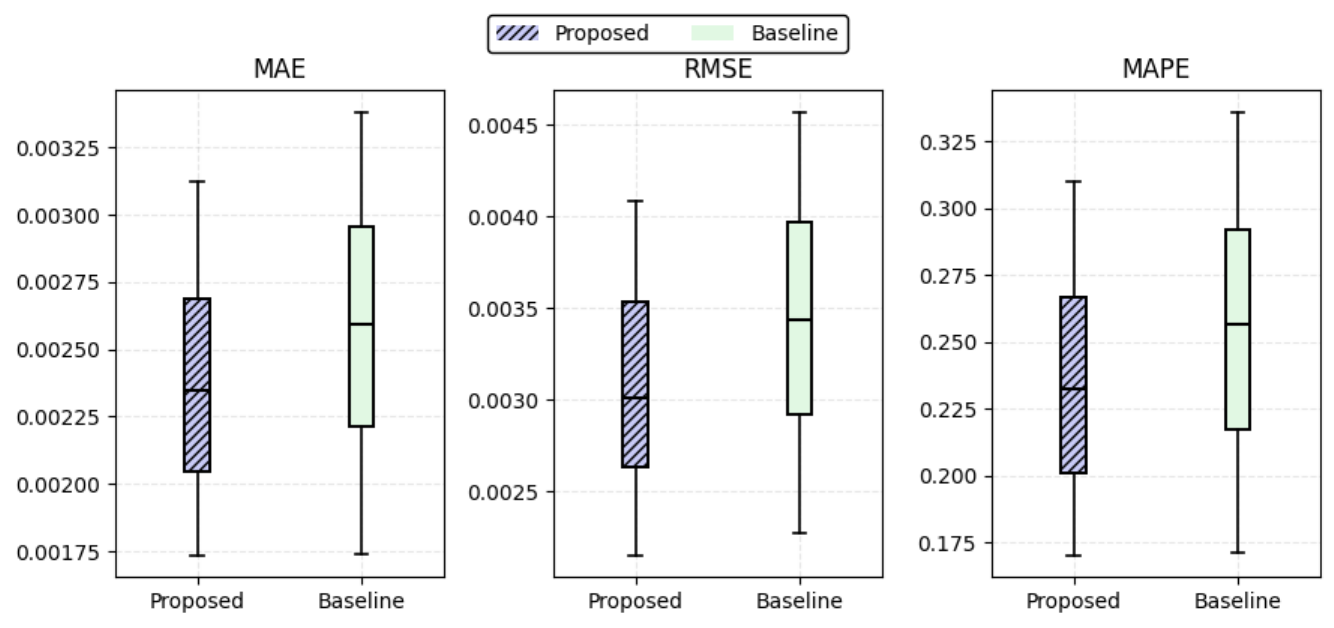}
    \caption{Performance comparison between the proposed and baseline methods for voltage recovery.}
    \label{fig:box plot v}
\end{figure}

The proposed method is also compared against the baselines described in the case-study section. Table~\ref{tab:mae volt baseline} reports the MAE values for both methods, showing consistent improvement across all clusters. Cluster~1 exhibits the smallest improvement, while Cluster~4 shows the largest improvement. The baselines suffer reduced accuracy when many intracluster sensors are assigned to the same LDST, resulting in correlated data losses. On average, the proposed communication-aware framework provides a 7.33\% improvement in MAE across all clusters. A comparison of all three metrics (MAE, RMSE, and MAPE) is provided in Fig.~\ref{fig:box plot v}, further confirming the superiority of the proposed recovery method. All performance metrics are evaluated only on the missing entries.

\begin{figure}[t]
    \centering
    \includegraphics[width=0.48\textwidth]{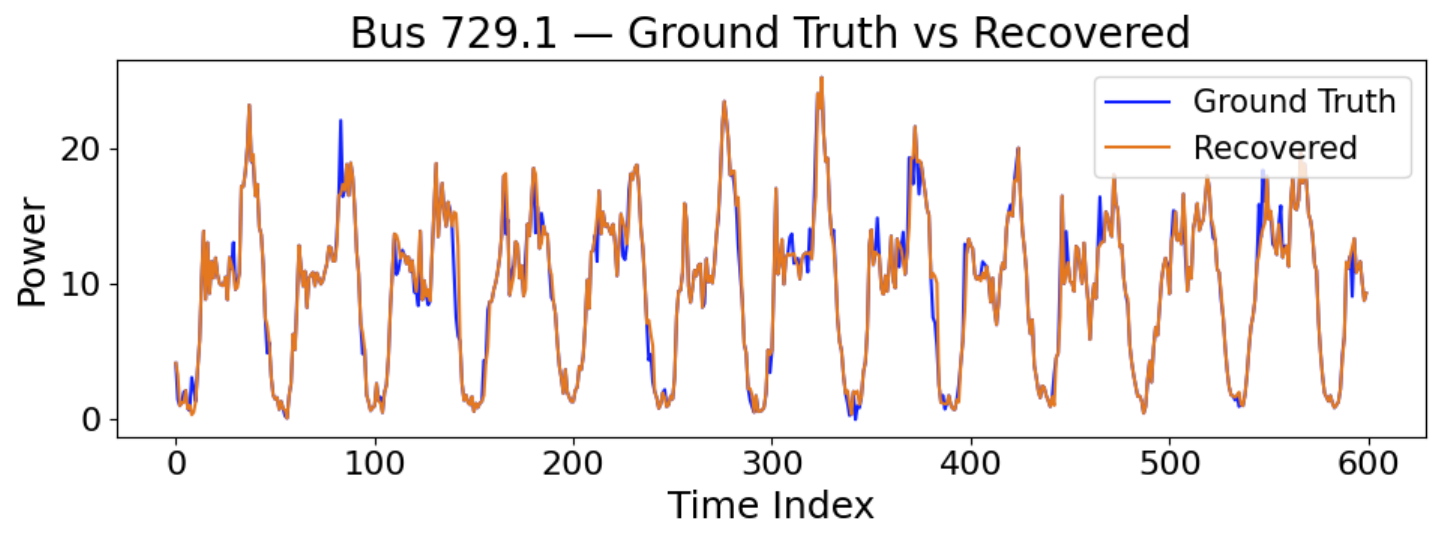}
\caption{Ground-truth versus recovered power injection measurements for a single node.}
    \label{fig: power overlap}
\end{figure}

\textbf{2) Power Injection Measurement Recovery.} 
The recovery performance is also evaluated for nodal power injection measurements. Only load buses are considered, since zero-injection buses do not require measurement recovery. Similar to the voltage measurement case, five clusters are formed for the power injection sensors, and corresponding LDSTs are constructed to collect their data. Monte Carlo simulation is then used to generate missing entries, which are subsequently recovered using OSVT.
For power injection data, the optimal Page matrix configuration was found to be 15 time windows with $L = 8$. To illustrate the recovery quality, the ground-truth and reconstructed injection measurements for a representative node are shown in Fig.~\ref{fig: power overlap}. The MAE comparison against the baselines is provided in Table~\ref{tab:mae_pi_baseline}. Consistent with the voltage recovery results, the proposed method outperforms the baseline across all clusters, achieving an average improvement of 12.93\%. These results further confirm the effectiveness of the proposed communication-aware data recovery framework.



\begin{table}[t]
    \centering
\caption{MAE comparison between the proposed and baseline methods for real power injection recovery under multiple link failures.}
    \label{tab:mae_pi_baseline}
    \begin{tabular}{ccccc}
        \hline
        \textbf{\textit{K}} & \textbf{Proposed} & \textbf{Baseline1} & \textbf{Baseline2} & Improvement (\%) \\
        \hline
        1        & 2.046337 & 2.831438 & 3.34304 & \textbf{27.75} \\
        2        & 1.070395 & 1.165941 & 1.35936 & \textbf{8.20}  \\
        3        & 4.545302 & 4.945750 & 6.17862 & \textbf{8.11}  \\
        4        & 2.418528 & 2.640774 & 3.05106 & \textbf{8.40}  \\
        5        & 1.193225 & 1.437139 & 1.64118 & \textbf{16.99} \\
        \hline
        \textbf{Combined} & \textbf{2.315946} & \textbf{2.659312} & \textbf{3.19118} & \textbf{12.93} \\
        \hline
    \end{tabular}
\end{table}

\section{Conclusion}
This paper developed a communication-aware measurement recovery framework that explicitly integrates communication routing constraints with low-rank reconstruction for resilient distribution system monitoring. Sensors were first grouped into balanced clusters, after which LDSTs were constructed to ensure that only a limited fraction of intracluster sensors shared the same communication path. This design preserves sufficient measurement diversity within each cluster under link failures, enabling effective OSVT-based recovery. Case studies on the IEEE 37-node test feeder using real-world smart meter data demonstrated consistent improvements in recovering both voltage and power injection measurements. The proposed method achieved an average MAE improvement of 7.33\% for voltage recovery and 12.93\% for power injection recovery, outperforming the communication-agnostic baseline across all evaluation metrics.

Future work will explore recovery on a large-scale system for additional measurement types, unified LDST formation for heterogeneous sensors, the use of higher-granularity data, and dynamic clustering and LDST reconfiguration enabled by software-defined networking.
\bibliographystyle{IEEEtran}
\bibliography{myref}

\end{document}